\RequirePackage{fix-cm}
\documentclass{svjour3}                     
\smartqed  
\usepackage{graphicx}
\usepackage{bm}
\usepackage{txfonts}
\begin{document}

\title{Supercurrent generation by spin-twisting itinerant motion of electrons
}
\subtitle{ }


\author{Daichi Manabe and Hiroyasu Koizumi 
}


\institute{D. Manabe, H. Koizumi \at
              Center for Computational Sciences, University of Tsukuba,Tsukuba, Ibaraki 305-8577, Japan}           

\date{Received: date / Accepted: date}

\maketitle

\begin{abstract}
Superconductivity is a phenomena where an external feeding current flows through the system without voltage drop.
This is explained if an energy minimum exists under the current feeding boundary condition.
We have found such minima in the system exhibiting spin twisting itinerant motion of electrons with the Rashba spin-orbit interaction
due to constraints of the charge conservation and single-valued requirement of the wave function. The supercurrent is shown to have the London equation form that exhibits the flux quantum $h/2e$. 
\end{abstract}

\section{Introduction}

Superconductivity is a phenomenon in which externally supplied electric current flows through a system without voltage drop \cite{O1,O2}. 
This current flowing state is dissipationless, thus, thermodynamically stable.
At zero temperature, this will lead to the conclusion that this current flowing state is energetically stable one, thus, must be one of local energy minimum states.
 
In the BCS theory based understanding, the mechanism for the supercurrent generation is explained using the dc Josephson effect \cite{Josephson62}.
The Josephson effect occurs in a superconductor-insulator-superconductor junction
 (Josephson junction), where the insulator part is so thin that electrons can tunnel through it. Josephson considered the Cooper pair tunneling through it;
each superconductor in the junction is assumed to be characterized by the phase conjugate to the Cooper pair number, and predicted that the dissipationless current is generated by the difference of the phases of the two superconductors.
Then, regarding 
 superconductor as a collection of Josephson junctions, the supercurrent in a superconductor is explained as due to the gradient of the phase. 
 But this explanation does not guarantee the existence of  an energy minimum under the external current feeding condition.

Through the study of high temperature superconductivity in cuprates, a new supercurrent generation mechanism has been proposed by one of the present authors \cite{Koizumi2011,Hidekata2011,HKoizumi2013,HKoizumi2014}. The spin-twisting circular motion of electrons is shown to generate supercurrent. 
In the present work, we further consider this mechanism.
We will show that if the Rashba spin-orbit interaction exists, energy minima occur at nonzero feeding current.

The organization of the present work is following: In Section \ref{sec2}, we explain the model in this work. In Section \ref{sec3}, we present calculations that demonstrate the existence of stable current carrying states under nonzero external current feeding.
In Section \ref{sec4}, we show that supercurrent of the present theory exhibits the London equation with flux quantum $h/2e$.
Lastly, we conclude the present work in Section \ref{sec5}.

\section{A model Hamiltonian for realizing spin-twisting itinerant motion of electrons}
\label{sec2}
 
In this section, we explain the model that realizes the dissipationless current  carrying state under the external current feeding boundary condition.
It was originally developed to explain superconductivity in cuprates. 
It  takes into account the following experimental facts: 

 1) The parent compound of the cuprate superconductor is a Mott insulator. It  is well-described by the Hubbard model.

 2) Bulk-sensitive experiments indicate small polaron formation due to strong hole-lattice interaction \cite{Bianconi,Miyaki2008}. Note that this small polaron formation is suppressed in the surface region where an energy gap with d-wave pairing profile is observed, due to the absence of the charge layer that covers the CuO$_2$ plane and stabilizes the polaron. 
 
 \
 
The small polarons in the bulk CuO$_2$ planes give rise to the following two important effects:

 1) They create an antiferromagnetic exchange interaction between electrons across the hole occupied sites. Then, frustration in spins occur since another antiferromagnetic exchange interaction between electrons in the nearby copper sites arises form the Hubbard Hamiltonian, resulting the creation of spin-vortices around the hole occupied sites.
 
  2) They create an internal electric field with the component perpendicular to the CuO$_2$ plane around the hole occupied sites. This gives rise to a Rashba spin-orbit interaction when electric current exists; actually, the spin-vortices mentioned above creates spin-vortex-induced loop currents (SVILC)  generated as a whole system motion due to the single-valued constraint of the wave function\cite{HKoizumi2013,HKoizumi2014}; thus, the nonzero Rashba spin-orbit interaction occurs.  Since the electric current is in the CuO$_2$ plane and the internal electric field has a component perpendicular to it, the spin vortices with twisting components in the CuO$_2$ plane arise.
  
  \

From the above consideration, we have constructed the following model Hamiltonian for electrons in the two-dimensional square lattice for the bulk CuO$_2$ plane,
\begin{eqnarray}
 H
  &=&-t\sum_{\langle i,j \rangle_{1},\sigma}(c^\dagger_{i\sigma}c_{j\sigma}+c^\dagger_{j\sigma}
   c_{i\sigma})
  +U\sum_{j} c^{\dagger}_{j\uparrow} c_{j\uparrow} c^{\dagger}_{j \downarrow}c_{j\downarrow}
  +J_h \sum_{\langle i,j \rangle_{h}} \hat{\bf S}_{i}\cdot \hat{\bf S}_{j}+H_{\rm so}
  \label{Ham}
  \end{eqnarray}
   Here, oxygens between nearest neighbor coppers are not explicitly taken into account, and $c^{\dagger}_{j \sigma}$ and  $c_{j \sigma}$ are the creation and annihilation operators of electrons at the $j$th site with the $z$-axis projection of electron spin $\sigma$, respectively. The first two terms come form the Hubbard Hamiltonian; $\langle i, j \rangle_1$ indicates the nearest neighbor site pairs. The parameter $t$ is the transfer integral and it is the units of energy in this work, and  $U$ is the on-site Coulomb parameter; we will adopt $U=8t$ in the later calculation, a typical value for the cuprate. 
 The third term describes the antiferromagnetic exchange interaction across the hole occupied sites, and $J_h$ is the coupling constant for it; we adopt $J_h=0.25t$ (Note that the value for the nearest neighbor antiferromagnetic exchange interaction parameter is $J=4t^2/U=0.5t$). 
 For the sum over $i$ and $j$, hole occupied sites are excluded assuming that they are immobile due to the small polaron formation. The sum over $\langle i, j \rangle_h$ is the sum over the pairs across the hole occupied sites, including the right angle directions. The spin operator at the $j$th site $\hat{\bf S}_j$ is given by 
 \begin{eqnarray}
\hat{\bf S}_j={ 1 \over 2} \sum_{\sigma, \sigma' } c_{j \sigma}^{\dagger} {\bm \sigma}_{\sigma \sigma'} c_{j \sigma'}.
\end{eqnarray}
where ${\bm \sigma}$ is the vector of Pauli matrices.
 
  The fourth term $H_{\rm so}$ is that for the Rashba spin-orbit interaction,
  \begin{eqnarray}
  H_{\rm so}
 &=&\! \lambda\sum_{h} \Big[ c^{\dagger}_{h\!+\!y \downarrow}c_{h\!-\!x \uparrow}\!-\!c^{\dagger}_{h\!+\!y \uparrow}c_{h\!-\!x \downarrow}\!+\!i(c^{\dagger}_{h\!+\!y\downarrow}c_{h\!-\!x\uparrow}\!+\!c^{\dagger}_{h\!+\!y\uparrow} c_{h\!-\!x\downarrow})
\nonumber
\\
\!&+&\! c^{\dagger}_{h\!+\!x \downarrow}c_{h\!-\!y \uparrow}\!-\!c^{\dagger}_{h\!+\!x \uparrow}c_{h\!-\!y \downarrow}\!+\!i(c^{\dagger}_{h\!+\!x\downarrow}c_{h\!-\!y\uparrow}\!+\!c^{\dagger}_{h\!+\!x\uparrow} c_{h\!-\!y\downarrow})
\nonumber
\\
\!&+&\!c^{\dagger}_{h-x\downarrow}c_{h-y\uparrow}-c^{\dagger}_{h-x\uparrow}c_{h-y\downarrow}+i(c^{\dagger}_{h-x\downarrow}c_{h-y\uparrow}+c^{\dagger}_{h-x\uparrow}c_{h-y\downarrow})
\nonumber
\\
\!&+&\!c^{\dagger}_{h+y\downarrow}c_{h+x\uparrow}-c^{\dagger}_{h+y\uparrow}c_{h+x\downarrow}+i(c^{\dagger}_{h+y\downarrow}c_{h+x\uparrow}+c^{\dagger}_{h+y\uparrow}c_{h+x\downarrow})
\nonumber
\\
&+& {\rm h.c.}\Big]
  \end{eqnarray}
 where $h$ describes the hole occupied sites \cite{Rashba2013,Koizumi2017}. We restrict that holes do not come nearby due to the Coulomb repulsion. $h+ x$ ( $h-x$ ) are nearest neighbor sites of $h$ in the $x$ direction (in the $-x$ direction); and  $h+y$ ( $h-y$ ) are nearest neighbor sites of $h$ in the $y$ direction (in the $-y$ direction). Here, we have assumed that the Rashba interaction exists only around the holes with the internal electric field in the direction perpendicular to the CuO$_2$ plane; the electric field is generated by the positive charge of the hole and the compensating charge due to dopant atoms in the charge reservoir layer, and the major component of it is assumed in the direction perpendicular to the CuO$_2$ plane since it is expected that the doped hole is more stable in the position of the CuO$_2$ plane close to the dopant atoms (for example, Sr for La$_{2-x}$Sr$_x$CuO$_4$). The internal electric field created this way will exist even for the cuprates whose parent compounds have a mirror symmetry with respect to the CuO$_2$ plane since the substituted atoms break the local symmetry around the small polaron. However,  the direction of the internal electric field may change either upwards or downwards, locally, with respect to the CuO$_2$ plane. In the present work, we only consider the case where the direction of the internal electric field around the holes is upwards throughout the sample.
$ \lambda$ is the parameter for the Rashba interaction; we adopt $\lambda=-0.02t$ for the most of the calculations below.

Since the hole occupied sites are excluded from the accessible sites,
the electron system is in the situation where the number of electrons and that of the accessible sites are equal.
We call this situation, the {\em effectively-half filled situation} (EHFS). The ordinary current generation by single-particle excitations is not effective due to the large energy gap of $U$ between the occupied lower band and empty upper band. The current we concern is the spin-vortex-induced loop current (SVILC)  generated as a whole system motion due to the single-valued constraint of the wave function\cite{HKoizumi2013,HKoizumi2014}.
 
The many-body Hamiltonian in Eq.~(\ref{Ham}) is too difficult to solve as it is; therefore, we use the following mean field version,
\begin{eqnarray}
H^{HF}_{EHFS}&=&-t\sum_{\langle i,j \rangle_{1},\sigma}(c^\dagger_{i\sigma}c_{j\sigma}+c^\dagger_{j\sigma}
   c_{i\sigma})
   \nonumber
   \\
  &+&U\sum_{j}\Big[ (\frac{1}{2}-S^z_j)c^\dagger_{j\uparrow}c_{j\uparrow}+
   (\frac{1}{2}+S^z_j)c^\dagger_{j\downarrow}c_{j\downarrow}
   \nonumber
   \\
   &-&(S^x_j-iS^y_j)c^\dagger_{j\uparrow}c_{j\downarrow}
   -(S^x_j+iS^y_j)c^\dagger_{j\downarrow}c_{j\uparrow}-{2 \over 3}{\bf S}^2 \Big]
   \nonumber
   \\
   &+&J_h\sum_{\langle i,j \rangle_{h}} \left( {\bf S}_{i}\cdot\hat{{\bf S}}_{j}+{\bf S}_{j}\cdot\hat{{\bf S}}_{i} -{\bf S}_{j}\cdot{{\bf S}}_{i}\right)
   +H_{\rm so}
   \label{hhf}
\end{eqnarray}
and  ${S}^{x}_j, S^{y}_j$ and $S^{z}_j$ are expectation values of the components of $\hat{\bf S}_j$ calculated as
\begin{eqnarray}
 {S}^{x}_j&=&\frac{1}{2}\langle c^{\dagger}_{j\uparrow} c_{j\downarrow}+c^{\dagger}_{j\downarrow} c_{j\uparrow}\rangle
 =S_j\cos\xi_j\sin\zeta_j
 \label{s3eq1}
 \\
 {S}^{y}_j&=&\frac{i}{2}\langle-c^{\dagger}_{j\uparrow}\ c_{j\downarrow}+c^{\dagger}_{j\downarrow} c_{j\uparrow}\rangle
 =S_j\sin\xi_j\sin\zeta_j
 \label{s3eq2}
 \\
 {S}^{z}_j&=&\frac{1}{2}\langle c^{\dagger}_{j\uparrow} c_{j\uparrow}-c^{\dagger}_{j\downarrow} c_{j\downarrow}\rangle
 =S_j\cos\zeta_j
 \label{s3eq3}
\end{eqnarray}
with $\langle \hat{O} \rangle$ denoting the expectation value of the operator $\hat{O}$. 

Through the self-consistent calculation using $H^{HF}_{EHFS}$, we obtain the following Hartree-Fock orbitals;
 \begin{eqnarray}
 |\tilde{\gamma_k} \rangle=\sum_{j} [ \tilde{D}_{j \uparrow}^{\gamma_k} c^{\dagger}_{j \uparrow}+\tilde{D}_{j \downarrow}^{\gamma_k} c^{\dagger}_{j \downarrow}]  | {\rm vac} \rangle.
 \end{eqnarray}
 We employ the Car-Parrinello method with a friction term to obtain $\tilde{D}^{\gamma}_{j \sigma}$'s.
 
 It is worth noting that $H_{\rm so}$ does not contribute at all for the evaluation of $\tilde{D}^{\gamma}_{j \sigma}$'s since the energy minimization procedure gives a currentless state.
 The Hartree-Fock orbitals $\{ |\tilde{\gamma}_k \rangle \}$ are so constructed to satisfy the orthonormal condition
 \begin{eqnarray}
\langle\tilde{\gamma}_j  |\tilde{\gamma}_k \rangle =\delta_{j k}.
 \end{eqnarray}
 
 A tentative total wave function $|\tilde{\Psi} \rangle$ is constructed as the Slater determinant of the occupied $|\tilde{\gamma}_k \rangle $'s.
Then, the expectation value of the spin components $S^x_j$ and $S^y_j$ are calculated using Eq.~(\ref{s3eq3}).
The angular value $\xi_j$ is obtained from $S^x_j$ and $S^y_j$; note that there is an ambiguity of an integral multiple of $2\pi$. 
We choose a particular branch of the multi-valued $\xi_j$. Note that the antiferromagnetic background from the Hubbard Hamiltonian given by $\xi^0_{j}=\pi (j_x+j_y)$ exists, where $(j_x, j_y)$ is the $xy$ coordinates of the $j$th site taking the lattice constant $a=1$.

We separate the antiferromagnetic contribution from $\xi$ and introduce angular variable $\eta$, 
 \begin{eqnarray}
 \eta_j=\xi_j-\pi (j_x+j_y)
 \label{eta}
 \end{eqnarray}
  where $\eta_j$ is $\eta$ at the $j$th site.
 We take the branch of $\eta_j$ that satisfies the difference of value from the nearest neighbor site $k$ is in the range,
\begin{eqnarray}
-\pi \le \eta_{j} -\eta_k < \pi
\end{eqnarray}

From ($\eta_j-\eta_k$)'s, we construct  ($\xi_j-\xi_k$)'s . After ($\xi_j-\xi_k$)'s are obtained, we rebuild $\xi$ from them. 
The process is as follows: first, we pick a value for the initial $\xi_1$  (say $\xi_1=0$). 
After fixing the value of $\xi_1$, we calculate $\xi_2$ by $\xi_{2} = \xi_1 + (\xi_{2} -\xi_1)$, where the site $2$ is connected to the site $1$ by a nearest neighbor bond.
 The step where value $\xi_{j}$ is derived from the already evaluated value of $\xi_k$ is given by
\begin{eqnarray}
\xi_{j} = \xi_k + (\xi_{j} -\xi_k)
\end{eqnarray}
where the sites ${j}$ and $k$ are connected by a bond in the path for the rebuilding of $\xi$. 
This process is continued until values at all accessible sites are evaluated once and only once.
By this rebuilding process, a single path is constructed from the site $1$ to other sites $k\neq 1$. 
We denote it by $C_{1 \rightarrow k}$.
Then, the value $\xi_k$ is given as
\begin{eqnarray}
\xi_k \approx \xi_1+ \int_{C_{1 \rightarrow k}} \nabla \xi \cdot d{\bf r}
\label{rebuilxi}
\end{eqnarray}

The presence of spin-vortices are described by non-zero winding numbers, $w_{C_{\ell}}[\xi]$, where the winding number of $\xi$ for loop $C_{\ell}$ that encircles a hole is defined by
\begin{eqnarray}
w_{C_{\ell}}[\xi]={ 1 \over {2\pi}} \sum_{i=1}^{N_{\ell}} ( \xi_{C_{\ell}(i+1)} -\xi_{C_{\ell}(i)}) \approx { 1 \over {2\pi}} \oint_{C_{\ell}} \nabla \xi \cdot d{\bf r}
\label{wnumber}
\end{eqnarray}
where $N_{\ell}$ is the total number of sites on the loop $C_{\ell}$, and ${C_{\ell}(i)}$ is the $i$th site on it with the periodic condition ${C_{\ell}(N_{\ell}+1)=C_{\ell}(1)}$. 
Here, we anticipate the spin is polarized in the CuO$_2$ plane due to the Rashba interaction. 

The angle $\xi$ may have jump-of-values (integer multiple of $2\pi$) between bonds that are not used in the precess of evaluating its value. This jump-of-value causes
the multi-valuedness in $|\tilde{\gamma} \rangle$, since $|\tilde{\gamma} \rangle$ is actually expressed as
 \begin{eqnarray}
 |\tilde{\gamma} \rangle=\sum_{j} \left[ e^{-i{{\xi_j} \over 2}}{D}_{j \uparrow}^{\gamma} {c}^{\dagger}_{j \uparrow}+ e^{i{{\xi_j} \over 2}}{D}_{j \downarrow}^{\gamma} {c}^{\dagger}_{j \downarrow} \right]  | {\rm vac} \rangle
  \label{tabhenkan2}
 \end{eqnarray}
 It contains factors $e^{\pm i{{\xi_j} \over 2}}$ that become multi-valued in the presence of the spin-vortices.
 
Due to the multi-valuedness of $ |\tilde{\gamma} \rangle$, $|\tilde{\Psi} \rangle$ becomes multi-valued.
On the other hand, the exact total wave function must be single-valued as a function of electron coordinates. We remedy this discrepancy by adding a phase factor that 
compensates the multi-valuedness of the basis $\{ |\tilde{\gamma} \rangle \}$. Namely, 
we construct the single-valued basis $\{ |{\gamma} \rangle \}$ given by
  \begin{eqnarray}
 |\gamma \rangle& =&\sum_{j} e^{\!-\!i { {\chi_j } \over 2}} [e^{-i{{\xi_j} \over 2}}D_{j \uparrow}^{\gamma} {c}^{\dagger}_{j \uparrow}\!+\!e^{i{{\xi_j} \over 2}}D_{j \downarrow}^{\gamma} {c}^{\dagger}_{j \downarrow}] | {\rm vac} \rangle
\label{Cwave}
 \end{eqnarray}
 and obtain the single-valued total wave function $|{\Psi} \rangle$ as the Slater determinant of the occupied $|{\gamma}_k \rangle $'s.
 
 Before obtaining $\chi$, we need to evaluate ${D}^{\gamma}_{j \sigma}$'s that are compatible with the rebuilt $\xi$ obtained using $C_{1 \rightarrow k}$'s, where
 the compatible ${D}^{\gamma}_{j \sigma}$'s means they are obtained for $\xi$ whose jump-of-value locations are known.
  It is important that $\chi$ has the same jump-of-value locations.
  In order to obtain ${D}^{\gamma}_{j \sigma}$'s, we diagonalize the following Hamiltonian one time using ${\bf S}_j$ and $n_j$ obtained above;
\begin{eqnarray}
\tilde{H}^{HF}_{EHFS}&=&-t\sum_{\langle i,j \rangle_{1}}\left(e^{i {1 \over2}(\xi_i -\xi_j)} \tilde{c}^\dagger_{i\uparrow} \tilde{c}_{j\uparrow}+
e^{-i {1 \over2}(\xi_i -\xi_j)} \tilde{c}^\dagger_{i\downarrow} \tilde{c}_{j\downarrow} + {\rm H.c.} \right)
   \nonumber
   \\
  &+&U\sum_{j}\Big[ (\frac{1}{2}-S^z_j)\tilde{c}^\dagger_{j\uparrow} \tilde{c}_{j\uparrow}+
   (\frac{1}{2}+S^z_j) \tilde{c}^\dagger_{j\downarrow} \tilde{c}_{j\downarrow}
   \nonumber
   \\
   &-&(S^x_j-iS^y_j)\tilde{c}^\dagger_{j\uparrow}\tilde{c}_{j\downarrow}
   -(S^x_j+iS^y_j)\tilde{c}^\dagger_{j\downarrow} \tilde{c}_{j\uparrow}\Big]
   \nonumber
   \\
   &+&J_h\sum_{\langle i,j \rangle_{h}} \left( {\bf S}_{i}\cdot\hat{{\bf S}}_{j}+{\bf S}_{j}\cdot\hat{{\bf S}}\right)
   \label{tildhhf}
\end{eqnarray}
where
\begin{eqnarray}
\tilde{c}^{\dagger}_{j\uparrow}= {c}^{\dagger}_{j\uparrow} e^{-i {1 \over2} \xi_i}, \  \tilde{c}_{j\uparrow}= {c}_{j\uparrow} e^{i {1 \over2} \xi_i}, \ 
 \tilde{c}^{\dagger}_{j\downarrow}= {c}^{\dagger}_{j\downarrow}e^{i {1 \over2} \xi_i}, \  \tilde{c}_{j\downarrow}= {c}_{j\downarrow}e^{-i {1 \over2} \xi_i}
\end{eqnarray}
Now, the single-particle wave functions $|\tilde{\gamma} \rangle$ is given by
\begin{eqnarray}
 |\tilde{\gamma} \rangle=\sum_{j} \left[{D}_{j \uparrow}^{\gamma} \tilde{c}^{\dagger}_{j \uparrow}+ {D}_{j \downarrow}^{\gamma} \tilde{c}^{\dagger}_{j \downarrow} \right]  | {\rm vac} \rangle
 \end{eqnarray}

When spin-vortices are present, phase factors $e^{\pm i\frac{\xi_j}{2}}$ in Eq.~(\ref{Cwave}), become multi-valued with respect to the coordinate since
$\xi_j$ has ambiguity of adding an integral multiple of $2\pi$.
 To restore the single-valuedness, the phase $\chi$ satisfies the following condition,
 \begin{eqnarray}
w_{C_{\ell}}[\xi]+ w_{C_{\ell}}[\chi] = \mbox{even number}  \mbox{  for any loop $C_{\ell}$}
\label{windcond}
\end{eqnarray}

The angular variable $\chi$ for the ground state is obtained by minimizing the total energy by imposing the above constraint.
We obtain $(\chi_k-\chi_j)$'s by minimizing the following functional 
\begin{eqnarray}
F[\nabla \chi]=E[\nabla \chi]+\sum_{\ell=1}^{N_{\rm loop}} { {\lambda_{\ell}}}\left(  \oint_{C_\ell} \nabla \chi \cdot d {\bf r}-2 \pi w_{C_{\ell}}[\chi] \right), 
\label{functional}
\end{eqnarray}
where 
\begin{eqnarray}
E[\nabla \chi]=\langle {\Psi} | H^{HF}_{EHFS} |{\Psi} \rangle
\label{energyf}
\end{eqnarray}
 $\lambda_{\ell}$'s are Lagrange multipliers, and $\{ C_1, \cdots, C_{N_{\rm loop}} \}$ are boundaries of plaques of the lattice, where $N_{\rm loop}$ is equal to the
number of plaques of the lattice.
  
We take the branch of $\chi_j$ that satisfies the difference of value from the nearest neighbor site $k$ is in the range,
\begin{eqnarray}
-\pi \le \chi_{j} -\chi_k < \pi
\end{eqnarray}
 
We rebuild $\chi$ from  $(\chi_k-\chi_j)$'s in a similar manner as $\xi$ is rebuilt. Thus, $\chi_k$ is given by
\begin{eqnarray}
\chi_k \approx \chi_1+ \int_{C_{1 \rightarrow k}} \nabla \chi \cdot d{\bf r}
\label{rebuilchi}
\end{eqnarray}

$(\chi_k-\chi_j)$'s are obtained as solutions of the following system of equations;
\begin{eqnarray}
{{\delta F[\nabla \chi]} \over {\delta \nabla \chi}}={{\delta E[\nabla \chi]} \over {\delta \nabla \chi}}+\sum_{\ell=1}^{N_{\rm loop}} { {\lambda_{\ell}}} {{\delta } \over {\delta \nabla \chi}} \oint_{C_\ell} \nabla \chi \cdot d {\bf r}&=&0
\label{Feq1}
\\
 \oint_{C_\ell} \nabla \chi \cdot d {\bf r}&=&2 \pi w_{C_{\ell}}[\chi]
 \label{Feq2}
\end{eqnarray}
 A set of parameters $\{ w_{C_{\ell}}[\chi] \}$ must be supplied as part of boundary conditions. The number of them is $N_{\rm loop}$, which is equal to the number of 
 $\{ \lambda_{\ell} \}$ to be evaluated.

In the discrete lattice, Eqs.~(\ref{Feq1}) and (\ref{Feq2}) are given by
\begin{eqnarray}
{{\partial  E( \{ \tau_{k \leftarrow j} \}) } \over {\partial \tau_{k \leftarrow j} }}+\sum_{\ell=1}^{N_{\rm loop}}  { {\lambda_{\ell}}} 
{{\partial } \over {\partial \tau_{k \leftarrow j}}} \sum_{k \leftarrow j} L_{k \leftarrow j}^{\ell}\tau_{k \leftarrow j} &=&0
\label{Feq1b}
\\
 \sum_{k \leftarrow j} L_{k \leftarrow j}^{\ell}\tau_{k \leftarrow j} &=& 2 \pi w_{C_{\ell}}[\chi]
 \label{Feq2b}
\end{eqnarray}
where the sum is taken over the bonds ${k \leftarrow j}$, $\tau_{ j \leftarrow i}$ is the difference of $\chi$ for the bond $\{k \leftarrow j \}$
\begin{eqnarray}
\tau_{ j \leftarrow i}=\chi_j -\chi_i   
\end{eqnarray}
and $L_{k \leftarrow j}^{\ell}$ is defined as
\begin{eqnarray}
L_{k \leftarrow j}^{\ell} =
\left\{
\begin{array}{cl}
-1 & \mbox{ if  $ j \leftarrow i$ exists in $C_{\ell}$ in the clockwise direction}
\\
1 & \mbox{ if  $ j \leftarrow i$ exists in $C_{\ell}$ in the counterclockwise direction}
\\
0 & \mbox{ if  $ j \leftarrow i$ does not exist in $C_{\ell}$}
\end{array}
\right.
\end{eqnarray}
The number of equations in Eqs.~(\ref{Feq1b}) and (\ref{Feq2b}) is
(the number of bonds)+(the number of plaques) which is equal to the number of unknowns $\{ \tau_{ j \leftarrow i} \}$ and $\{ \lambda^{\ell} \}$.

The external current boundary condition is imposed by adding external loops to functional $F$ as
\begin{eqnarray}
F[\nabla \chi ]=E[\nabla \chi]+\sum_{\ell=1}^{N_{\rm loop}} { {\lambda_{\ell}}}\left(  \oint_{C_\ell} \nabla \chi \cdot d {\bf r}-2 \pi w_{C_{\ell}}[\chi] \right)
+\sum_{\ell=1}^{N^{\rm EX}_{\rm loop}} { {\lambda^{\rm EX}_{\ell}}} \oint_{C^{\rm EX}_\ell} \nabla \chi \cdot d {\bf r}
\label{FunctionalF2}
\end{eqnarray}
where $C^{\rm EX}_{\ell}$ is a external loop that connects a site in the lattice system to another site in the lattice; one of them is the site for flow-in and the other is for
flow-out of the external current. Note that $\lambda^{\rm EX}_{\ell}$ is not a Lagrange multiplier; it is determined by the direction and magnitude of the external current 
through $C^{\rm EX}_\ell$ as boundary conditions.

For the case with external loops, the equations in Eq.~(\ref{Feq1}) become
\begin{eqnarray}
{{\delta E[\nabla \chi]} \over {\delta \nabla \chi}}+\sum_{\ell=1}^{N_{\rm loop}} { {\lambda_{\ell}}} {{\delta } \over {\delta \nabla \chi}} \oint_{C_\ell} \nabla \chi \cdot d {\bf r}+\sum_{\ell=1}^{N^{\rm EX}_{\rm loop}} { {\lambda^{\rm EX}_{\ell}}} {{\delta } \over {\delta \nabla \chi}} \oint_{C^{\rm EX}_\ell}\nabla \chi \cdot d {\bf r}=0
\end{eqnarray}

We can actually obtain $\tau_{ j \leftarrow i}$'s without obtaining ${ {\lambda_{\ell}}}$'s. This method is convenient when the feeding current is introduced and the Rashba spin-orbit interaction is included. In the following calculations, we adopt this method.
We shall explain this method more in detail below.

First, we note that the current density is given by
\begin{eqnarray}
{\bf j}={{2e} \over \hbar}  {{\delta E} \over {\delta \nabla \chi}}
\end{eqnarray}
For the lattice system, it is expressed as
\begin{eqnarray}
J_{ j \leftarrow i}= {{2e} \over \hbar}  {{\partial E} \over {\partial \tau_{ j \leftarrow i}}}
\end{eqnarray}
where $J_{ j \leftarrow i}$ is the current through the bond between sites $i$ and $j$ in the direction  $j \leftarrow i$.

Then, the conservation of charge at site $j$ is given by
\begin{eqnarray}
0=J^{\rm EX} _{ j} + \sum_i {{2e} \over \hbar}  {{\partial E} \over {\partial \tau_{ j \leftarrow i}}}
\label{Feq3}
\end{eqnarray}
where $J^{\rm EX} _{ j}$ is the external current that enters through site $j$. We use this in place of Eq.~(\ref{Feq1b}).

In order to impose conditions in Eq.~(\ref{Feq2b}), $\tau_{ j \leftarrow i}$ is split into a multi-valued part $\tau_{ j \leftarrow i}^0$ and 
single-valued part $f_{ j \leftarrow i}$ as
\begin{eqnarray}
\tau_{ j \leftarrow i}=\tau_{ j \leftarrow i}^0 + f_{ j \leftarrow i}
\end{eqnarray}
where $\tau_{ j \leftarrow i}^0$ satisfies the constraint in Eq.~(\ref{Feq2b})
\begin{eqnarray}
w_{C_{\ell}}[\chi]={ 1 \over {2\pi}} \sum_{i=1}^{N_{\ell}}  \tau^0_{{C_{\ell}(i+1)} \leftarrow {C_{\ell}(i)}}
\end{eqnarray}
and $f_{ j \leftarrow i}$ satisfies
\begin{eqnarray}
0={ 1 \over {2\pi}} \sum_{i=1}^{N_{\ell}}  f_{{C_{\ell}(i+1)} \leftarrow {C_{\ell}(i)}}
\end{eqnarray}

The equation (\ref{Feq3}) is used to obtain $f_{ j \leftarrow i}$'s in the present method. We employ an iterative improvement of the approximate solutions by using the linearized version of Eq.~(\ref{Feq3}) given by
\begin{eqnarray}
0 \approx J^{\rm EX} _{ j} + {{2e} \over \hbar} \sum_i  {{\partial E (\{ \tau_{ j \leftarrow i}^0 \}) } \over {\partial \tau_{ j \leftarrow i}}}+{{2e} \over \hbar} 
\sum_i  {{\partial^2 E (\{ \tau_{ j \leftarrow i}^0 \}) } \over {\partial (\tau_{ j \leftarrow i})^2}}f_{ j \leftarrow i}
\label{Feq4}
\end{eqnarray}
These equations are solved for $f_{ j \leftarrow i}$'s for given $\tau_{ j \leftarrow i}^0$'s. $\tau_{ j \leftarrow i}^0$'s are updated at each iteration as
\begin{eqnarray}
\tau_{ j \leftarrow i}^{ 0 \ 
New}=\tau_{ j \leftarrow i}^{ 0 \ Old}+f_{ j \leftarrow i}
\end{eqnarray}
where $\tau_{ j \leftarrow i}^{ 0 \ Old}$ is $\tau_{ j \leftarrow i}^{ 0}$ value that is used to obtain the current value of $f_{ j \leftarrow i}$; $\tau_{ j \leftarrow i}^{ 0 \ New}$ will be used to obtain the next $f_{ j \leftarrow i}$ value. The convergence is checked by the condition
\begin{eqnarray}
\left| J^{\rm EX} _{ j} + {{2e} \over \hbar} \sum_i  {{\partial E (\{ \tau_{ j \leftarrow i}^0 \}) } \over {\partial \tau_{ j \leftarrow i}}} \right| < \epsilon
\end{eqnarray}
where $\epsilon$ is a small number.

For the initial $\tau_{ j \leftarrow i}^{ 0 }$, we adopt the following,
\begin{eqnarray}
\tau_{ j \leftarrow i}^{ 0 \ init}=\sum_h  w_h \tan^{-1} {{ j_y- h_y} \over {j_x -h_x}}-\sum_h w_h \tan^{-1} {{ i_y- h_y} \over {i_x -h_x}}
\end{eqnarray}
where $(j_x,j_y)$ and $(i_x,i_y)$ are  coordinates of the sites $j$ and $i$, respectively, $h=(h_x,h_y)$ is the coordinate of the hole occupied site, and $w_h$ is the winding number of $\chi$ around the hole at $h$.

For the lattice system, the number of $\tau_{ j \leftarrow i}$ to be evaluated is equal to the number of bonds; i.e., the number of unknowns is equal to the number of bonds. The number of equations in Eq.~(\ref{Feq2b}) is equal to the number of plaques of the lattice; the number of equations from Eq.~(\ref{Feq3}) is equal to the number of sites-1, where '-1'  arises due to the fact that the conservation of the total charge is maintained in the calculation, thus, requiring the conservation of charge at all sites is redundant by one. 

The equality of the number of unknowns and the number of equations gives
\begin{eqnarray}
(\mbox{the number of bonds})=(\mbox{the number of plaques})+(\mbox{the number of sites}-1)
\nonumber
\\
\label{Euler1}
\end{eqnarray}

This agrees with the Euler's theorem for the two-dimensional lattice 
\begin{eqnarray}
(\mbox{the number of edges})=(\mbox{the number of faces})+(\mbox{the number of vertices}-1)
\nonumber
\\
\label{Euler2}
\end{eqnarray}
 It is interesting that the mathematical expression in Eq.~(\ref{Euler2}) can be interpreted in a physical way as those for the number of unknowns and the number of equations in Eqs.~(\ref{Feq2b}) and (\ref{Feq3}).

\section{Stable current carrying state under external current feeding boundary condition}
\label{sec3}

\begin{figure}
\includegraphics[scale=0.6]{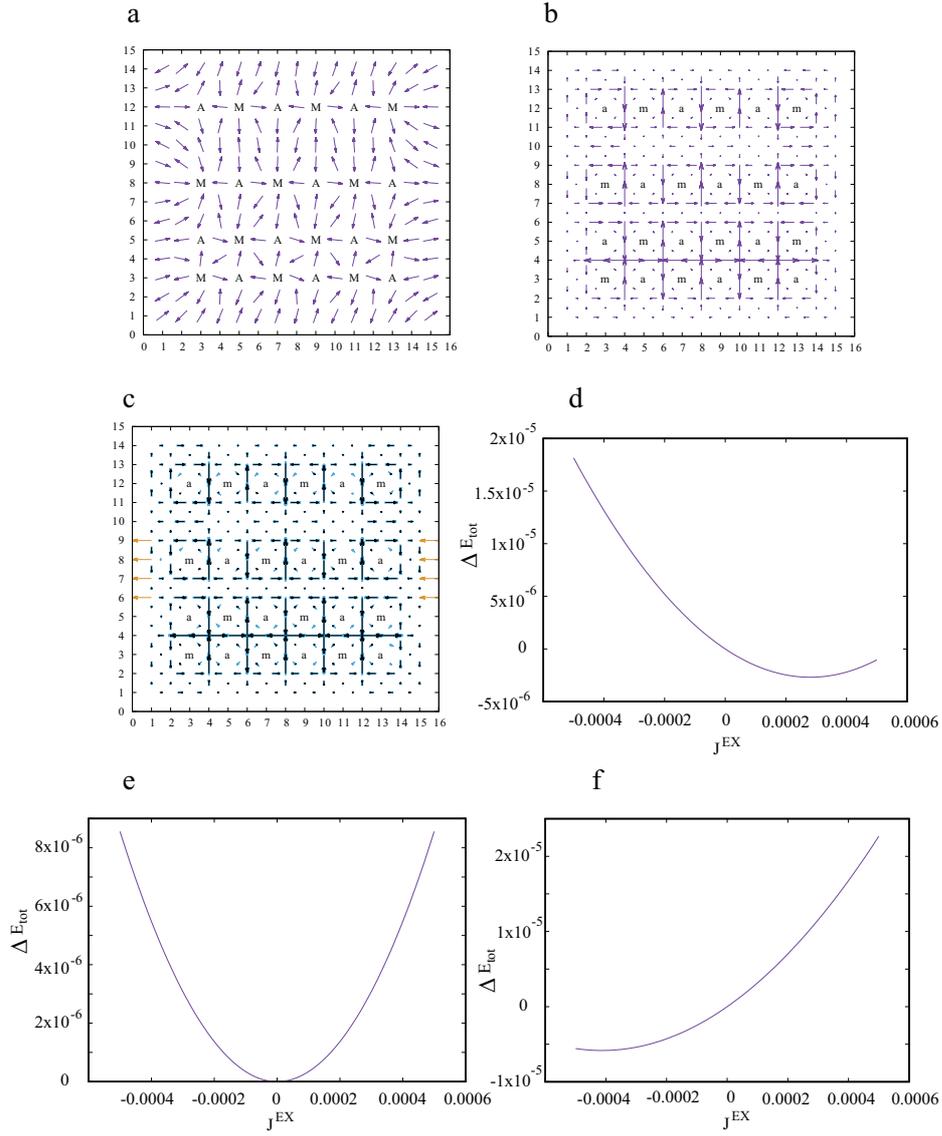}
\caption{ {Appearance of an energy minimum at nonzero external feeding current. The system is a $15 \times 15$ square lattice described by the Hamiltonian in Eq.~(\ref{hhf}). The Rashba spin-orbit interaction parameter used is $\lambda=-0.02t$ except e) and f).  a) The spin-texture. `M' and `A' indicate centers of spin vortices with winding numbers $+1$ and $-1$, respectively. b) The current
distribution. `m' and `a' indicate centers of loop currents with winding numbers $+1$ and $-1$, respectively. c) The current distribution (denoted by black arrows) when nonzero external current $J^{\rm EX}$ per site is fed; it flows-in at four sites (15,6)-(15,9) and flows-out at four sites (1,6)-(1,9) as indicated by brown arrows. In this figure, the value of  $J^{\rm EX}$ is set to the energy minimum value in d). Light-blue arrows show current for $J^{\rm EX}=0$. d) $\Delta E_{\rm tot}$ vs  $J^{\rm EX}$. $\Delta E_{\rm tot}$ is the difference of the total energy $E_{\rm tot}$ from its minimal value $E^{\rm tot}_{\rm min}$. e) The same as d) but with $\lambda=0$.  f) The same as d) but with $\lambda=0.02t$. }}
\label{Fig1}
\end{figure}

The system we consider is the $15 \times 15$ lattice with $24$ holes as shown in Fig.~\ref{Fig1}a. We only consider the cases where the winding numbers for $\xi$ (indicated by ``M'' or ``A'' in Fig.~\ref{Fig1}a) and those for $\chi$ (indicated by ``m'' or ``a'' in Fig.~\ref{Fig1}b) are the same for all spin-vortices and spin-vortex-induced loop currents. The minimal energy current pattern for a given spin texture is obtained by this winding number combination. The bonds taken into account are those of the square lattice excluding the ones connected to the hole occupied sites, and the second nearest bonds around  holes (four for each hole).  As a consequence, the plaques taken into account are those of the square lattice excluding the ones containing the hole occupied sites, triangles containing second nearest neighbor bonds around  holes (four for each hole), and
 squares containing second nearest neighbor bonds around  holes (one for each hole).
 
External current $J^{\rm EX}$ is fed as shown in Fig.~\ref{Fig1}c; in this figure, the minimal energy current pattern is superimposed on the
one without external current. The change of the current pattern by the external current feeding is most significant along the second nearest neighbor bonds around holes.
 In Fig.~\ref{Fig1}d, the total energy difference $\Delta E_{\rm tot}$ from the minimal value is plotted as a function of $J^{\rm EX}$. A minimum exists at around $J^{\rm EX}=0.0003$ in the units of $et/\hbar$. 
  The existence of this minimum means that nonzero feeding state is a stable one, namely, this feeding current is dissipationless.
  Thus, the current is the supercurrent, and the system is in the superconducting state. 
  
  The existence of the Rashba spin-orbit interaction is crucial to have the nonzero supercurrent.
 In Figs.~\ref{Fig1}e and \ref{Fig1}f, $\Delta E_{\rm tot}$ vs $J^{\rm EX}$ are depicted for the cases with the Rashba parameter $\lambda=0$ and $\lambda=0.02t$, respectively. The minimum disappears when $\lambda=0$. It is also notable that the stable current direction changes for the $\lambda=0.02t$ case compared with the the $\lambda=- 0.02t$ case in Fig.~\ref{Fig1}d.

\begin{figure} 
\includegraphics[scale=0.7]{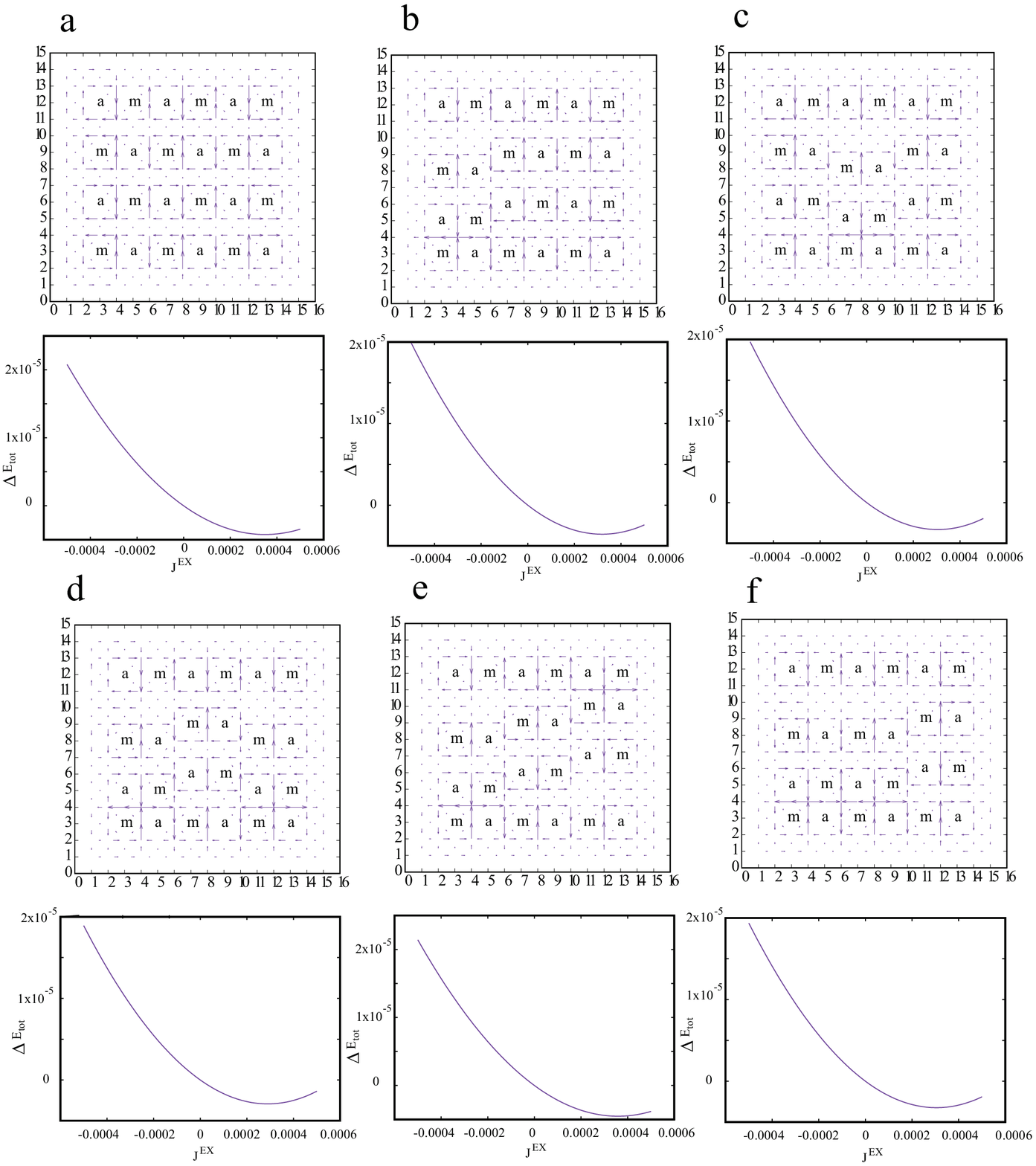}
\caption{ {Appearance of an energy minimum at nonzero external feeding current for various loop current patterns. In each figure, the current pattern is shown in the upper panel, and $\Delta E_{\rm tot}$ vs  $J^{\rm EX}$.} is shown in the lower panel. The underlying spin-vortex patterns are those obtained by replacing `m' by `M', and `a' by `A'.}
\label{Fig2}
 \end{figure}
 
 \begin{figure} 
\includegraphics[scale=0.8]{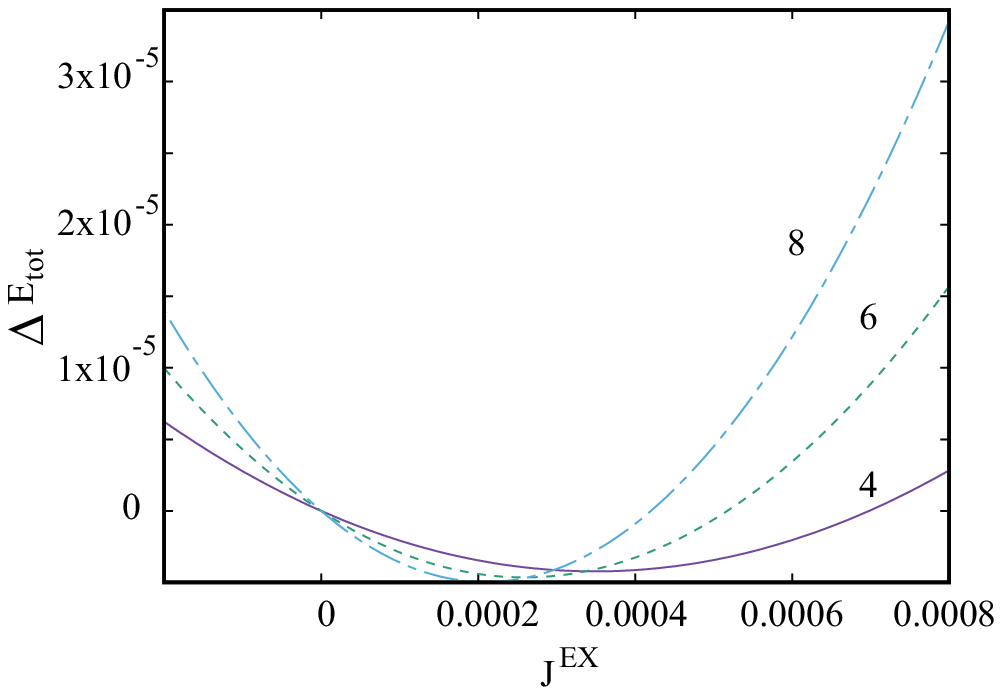}
\caption{ $\Delta E_{\rm tot}$ vs  $J^{\rm EX}$ for different number of current feeding sites. The system is the one shown in Fig.~\ref{Fig2}a. 
Sold line: $J^{\rm EX}$ flows-in at four sites (15,6)-(15,9) and flows-out at four sites (1,6)-(1,9); Doted line: $J^{\rm EX}$ flows-in at six sites (15,5)-(15,10) and flows-out at six sites (1,5)-(1,10); Dash-dot line: $J^{\rm EX}$ flows-in at eight sites (15,4)-(15,11) and flows-out at eights sites (1,4)-(1,11).}
\label{Fig3}
 \end{figure}
 
 In Fig.~\ref{Fig2}, $\Delta E_{\rm tot}$ vs  $J^{\rm EX}$ is depicted for  different number of current feeding sites. The energy minimum occurs at the similar value of $J^{\rm EX}$, thus, as the number of feeding sites is increased, the total current increases, roughly, proportionally.

\section{London equation and the flux quantization in $h/2e$}
\label{sec4}

\begin{figure}
\includegraphics[scale=0.5]{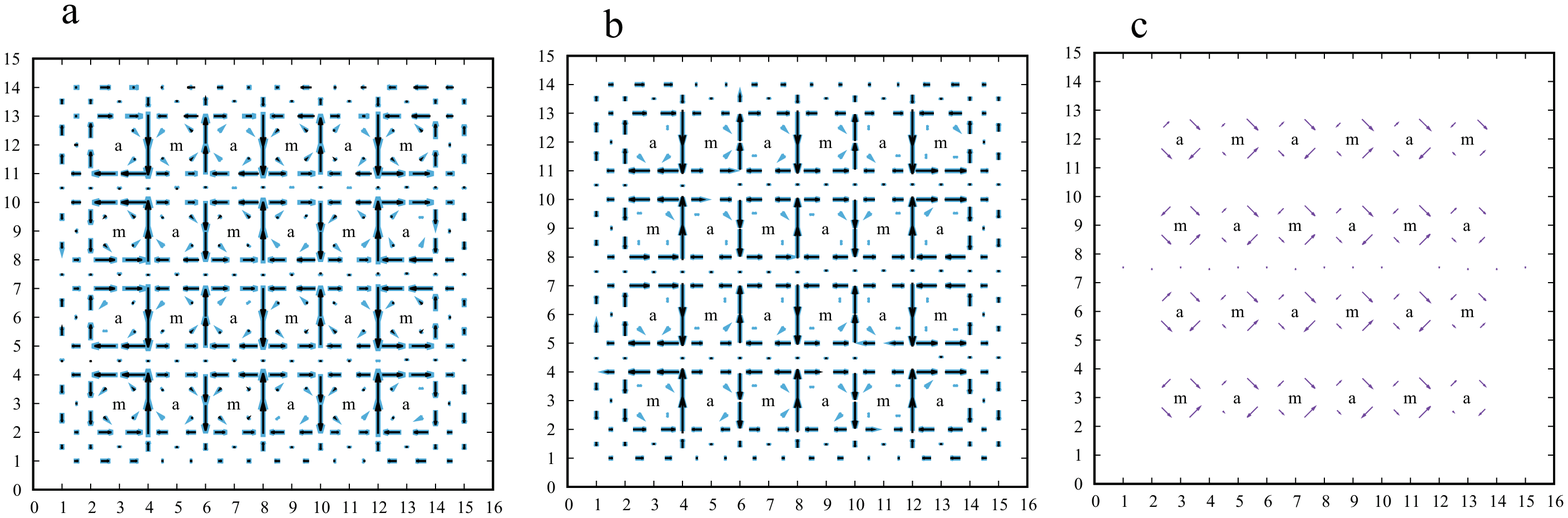}
\caption{ The current when a uniform magnetic field perpendicular to the plane is applied.  $B=0.01 \ \hbar /ea^2$. a) The current (denoted by black arrows) imposed on the current for ${\bf B}^{\rm em}=0$ (light-blue arrows, the same  in Fig.~\ref{Fig2}a). b). The part of the current linear in ${\bf A}^{\rm eff}$ (denoted by black arrows) imposed on the exact current (light-blue arrows, the same as black arrows in Fig.~\ref{Fig4}a). c) The part of the current constant in ${\bf A}^{\rm eff}$. Lengths of arrows are multiplied by 100 compared with b).}
\label{Fig4}
\end{figure}

Let us consider the situation where a magnetic field ${\bf B}^{\rm em}=\nabla \times {\bf A}^{\rm em}$ is applied.
Then, the energy functional  in Eq.~(\ref{energyf}) is modified as
\begin{eqnarray}
E[\nabla \chi] \rightarrow E \left[\nabla \chi -{{2e} \over \hbar}{\bf A}^{\rm em} \right]
\end{eqnarray}

This leads to replace $\tau_{ j \leftarrow i}$ in $E$ by
\begin{eqnarray}
u_{ j \leftarrow i}=\tau_{ j \leftarrow i}-{{2e} \over \hbar} \int^j_i{\bf A}^{\rm em}\cdot d{\bf r}
\end{eqnarray}
where integration is performed along the bond $ j \leftarrow i$.

The calculation can be done similarly from the initial value
\begin{eqnarray}
u_{ j \leftarrow i}^{0 \ init}=\tau_{ j \leftarrow i}^{ 0  \ init}-{{2e} \over \hbar} \int_{j \leftarrow i}{\bf A}^{\rm em}\cdot d{\bf r}
\end{eqnarray}

Note that during the evaluation process of $\nabla \chi$, the ambiguity in the gauge of ${\bf A}^{\rm em}$ is compensated, thus, making
\begin{eqnarray}
{\bf A}^{\rm eff}= {\bf A}^{\rm em}-{\hbar  \over {2e}} \nabla \chi
\end{eqnarray}
invariant with respect to the choice of the gauge in ${\bf A}^{\rm em}$. In other words, ${\bf A}^{\rm eff}$ is the gauge invariant vector potential in the material.

Now we apply a uniform magnetic field perpendicular to the lattice.
In the actual numerical calculations we have adopted
\begin{eqnarray}
{\bf A}^{\rm em}=
\left(
\begin{array}{c}
-By \\
0 \\
0
\end{array}
\right)
\end{eqnarray}
or 
\begin{eqnarray}
{\bf A}^{\rm em}=
\left(
\begin{array}{c}
0 \\
Bx\\
0
\end{array}
\right)
\end{eqnarray}

The resulting ${\bf A}^{\rm eff}$ obtained by the above different gauges are identical.
In Fig.~\ref{Fig4}a, the current pattern with the uniform magnetic field is depicted.

The current may be approximated as 
\begin{eqnarray}
J_{ j \leftarrow i}&=& {{2e} \over \hbar}  {{\partial E} \over {\partial u_{ j \leftarrow i}}}
\nonumber
\\
 &\approx& {{2e} \over \hbar}  {{\partial E( \{ 0 \} )} \over {\partial u_{ j \leftarrow i}} } +{{2e} \over \hbar}  {{\partial^2 E( \{ 0 \} )} \over {\partial (u_{ j \leftarrow i})^2}} u_{ j \leftarrow i}
 \nonumber
 \\
 &=& {{2e} \over \hbar}  {{\partial E( \{ 0 \} )} \over {\partial u_{ j \leftarrow i}} } -{{4e^2} \over \hbar^2}  {{\partial^2 E( \{ 0 \} )} \over {\partial (u_{ j \leftarrow i})^2}}
\int^j_i {\bf A}^{\rm eff}\cdot d{\bf r} 
\end{eqnarray}

In Fig.~\ref{Fig4}b, the linear term in ${\bf A}^{\rm eff}$ is depicted. It almost coincides with the exact result. A slight difference is found around the hole occupied sites.
This difference is essentially due to the constant term as shown in Fig.~\ref{Fig4}c. Similar results are obtained when the values (including sign) of $\lambda$ is changed. Thus, the linear relation between the current and ${\bf A}^{\rm eff}$ is a solid one.
 
 If we neglect the constant term and only keep the linear term, the current expression becomes the London equation \cite{London1948}.
 This equation yields the Meissner effect.
 If we take loop $C$ along the current zero region, i.e. along ${\bf A}^{\rm eff}= {\bf A}^{\rm em}-{\hbar  \over {2e}} \nabla \chi=0$, we have
 \begin{eqnarray}
\oint_C {\bf A}^{\rm em}\cdot d{\bf r} ={\hbar  \over {2e}} \oint_C \nabla \chi\cdot d{\bf r} ={h \over {2e}}w_C[\chi]
 \end{eqnarray}
 where $w_C[\chi]$ is the winding number. This shows the flux quantization in the unit $h/2e$.

\section{Concluding remarks}
\label{sec5}

In the present work, we have shown that energy minima exist under the current feeding boundary condition at nonzero feeding current in the system exhibiting the spin-twisting itinerant motion of electrons under the influence of the Rashba spin-orbit interaction.
This nonzero feeding current may be considered as the supercurrent measured experimentally, and the system may be regarded as in the superconducting state.

Actually, the dc Josephson effect explanation of the supercurrent generation encounters a difficulty in explaining the stability of the state under external current feeding
since the current zero state is energy minimal. Bloch claimed that although the success of the BCS theory it might not be the final theory because of the lack of obtaining current flowing energy minima \cite{Bloch1966}.
The present work indicates that the current flowing state becomes a minimum energy state due to the presence of constraints of the charge conservation and single-valued requirement of the wave function; namely, the constraints force the current to flow.
 The present work suggests that there is still room for fundamental improvement in the superconductivity theory.
 

\bibliographystyle{spphys}       

\end{document}